# Constructions for Difference Triangle Sets

Yeow Meng Chee and Charles J. Colbourn

*Abstract*—Difference triangle sets are useful in many practical problems of information transmission. This correspondence studies combinatorial and computational constructions for difference triangle sets having small scopes. Our algorithms have been used to produce difference triangle sets whose scopes are the best currently known.

## I. INTRODUCTION

An $(n, k)$-*difference triangle set*, or $(n, k)$-D$\Delta$S, is a set $\mathcal{X} = \{X_i | 1 \leq i \leq n\}$, where $X_i = \{a_{ij} | 0 \leq j \leq k\}$, for $1 \leq i \leq n$, are sets of integers called *blocks*, such that the differences $a_{ij} - a_{ij'}$, for $1 \leq i \leq n$ and $0 \leq j \neq j' \leq k$, are all distinct and nonzero. An $(n, k)$-D$\Delta$S is *normalized* if for $1 \leq i \leq n$, we have $0 = a_{i0} < a_{i1} < \cdots < a_{ik}$. All difference triangle sets considered in this correspondence are normalized. The *scope* of an $(n, k)$-D$\Delta$S, $\mathcal{X} = \{X_i | 1 \leq i \leq n\}$, is defined as

$$m(\mathcal{X}) = \max\left\{a \,\middle|\, a \in \bigcup_{i=1}^{n} X_i\right\}.$$

The smallest scope possible for an $(n, k)$-D$\Delta$S is

$$m(n, k) = \min\{m(\mathcal{X}) | \mathcal{X} \text{ is an } (n, k)\text{-D}\Delta\text{S}\}.$$

An $(n, k)$-D$\Delta$S $\mathcal{X}$ is *optimal* if $m(\mathcal{X}) = m(n, k)$. By counting differences, we easily obtain the *trivial lower bound*

$$m(n, k) \geq n\binom{k+1}{2}.$$

Better lower bounds can be found in the papers of Kløve [1], [2]. In particular, we have the following result [1, Theorem 2].

*Theorem 1 (Kløve):* For all $n$ and $k$

$$m(n, k) \geq n\left(k^2 - 2k\sqrt{k} + \frac{k + \sqrt{k}}{4}\right).$$

Given an $(n, k)$-D$\Delta$S $\mathcal{X}$, we can obtain an $(n-1, k)$-D$\Delta$S by omitting from $\mathcal{X}$ the block containing the largest element. This operation is called a *reduction*. The operation of omitting the largest element from each block of $\mathcal{X}$ is called *shortening*, and this gives an $(n, k-1)$-D$\Delta$S instead.

There is a restricted variant of difference triangle sets, called *regular perfect systems of difference sets* (see [3] and [4]), that is widely studied in combinatorial design theory. Let $c$ be a positive

Manuscript received November 7, 1995; revised June 24, 1996. This work was supported by the Natural Sciences and Engineering Research Council of Canada. The material in this correspondence was presented in part at the 27th Southeastern International Conference on Combinatorics, Graph Theory, and Computing, Baton Rouge, LA, February 1996. This work was performed while Y. M. Chee was with the Department of Computer Science, University of Waterloo, Waterloo, Ont. N2L 3G1, Canada, and while C. J. Colbourn was with the Department of Combinatorics and Optimization, University of Waterloo, Waterloo, Ont. N2L 3G1, Canada.

Y. M. Chee is with the Information Infrastructure Group, National Computer Board, Singapore 118253, Republic of Singapore.

C. J. Colbourn is with the Department of Computer Science and Electrical Engineering, University of Vermont, Burlington, VT 05405 USA.

Publisher Item Identifier S 0018-9448(97)03453-6.

integer. An $(n, k)$-D$\Delta$S $\mathcal{X} = \{X_i | 1 \leq i \leq n\}$, where $X_i = \{a_{ij} | 0 \leq j \leq k\}$, is a *regular* $(n, k+1, c)$-*PSDS* if

$$\bigcup_{i=1}^{n} \{a_{ij} - a_{ij'} | 0 \leq j' < j \leq k\}$$

$$= \left\{c, c+1, \cdots, c - 1 + n\binom{k+1}{2}\right\}.$$

The existence of a regular $(n, k+1, c)$-PSDS gives an $(n, k)$-D$\Delta$S of scope $c - 1 + n\binom{k+1}{2}$.

Difference triangle sets have a number of interesting applications in data communications (see [2] and [5]). For all of these applications, difference triangle sets with small scopes are desirable. Hence, the determination of $m(n, k)$ is of importance. Unfortunately, this is a rather difficult problem. The special case of determining $m(1, k)$ is the well-known problem of finding Golomb rulers, which has continued to resist many attacks. Only recently was it shown that $m(1, 18) = 246$ [6]. In general, except for the following result (see, for example, [5]) on $m(n, 1), m(n, 2)$, and $m(n, 3)$, only finitely many values of $m(n, k)$ are known.

*Theorem 2:* For $n \geq 1$, we have $m(n, 1) = n$ and

$$m(n, 2) = \begin{cases} 3n, & \text{if } n \equiv 0 \text{ or } 1 \pmod 4 \\ 3n + 1, & \text{if } n \equiv 2 \text{ or } 3 \pmod 4. \end{cases}$$

There are infinitely many values of $n$ for which $m(n, 3) = 6n$.

The value of $m(n, 1)$ is trivial. The value of $m(n, 2)$ is folklore and is derived from the existence of Skolem and Langford sequences [7], [8]. The result on $m(n, 3)$ follows from the results of Kotzig and Turgeon [9], and Rogers [10] on the existence of regular $(n, 4, 1)$-PSDS. The following conjecture was made by Bermond [11].

*Conjecture 1 (Bermond):* For every $n \geq 4$, we have $m(n, 3) = 6n$.

Bermond's conjecture has been verified for $4 \leq n \leq 22$ [12]. For each $k \in \{1, 2, 3\}$, the above results indicate that there exists an $n$ such that $m(n, k)$ meets the trivial lower bound. However, this phenomenon cannot persist, as it was shown in [13], [14] that $m(n, k) = n\binom{k+1}{2}$ only if $k \leq 3$, or $k = 4$ and $n$ is an even integer at least six.

The establishment of good bounds for $m(n, k)$ is, therefore, of interest.

Our concern in this correspondence is on the constructive aspects of difference triangle sets. The aim is to provide combinatorial as well as algorithmic constructions for difference triangle sets of small scope, thereby improving some of the existing upper bounds on $m(n, k)$. More information on difference triangle sets can be found in [1], [2], [5], [15], and [16].

## II. COMBINATORIAL ASYMPTOTICS

In this section, if $f$ and $g$ are two nonnegative functions, we use the notation $f \ll g$ to mean that there is an absolute constant $C$ such that $f \leq Cg$.

A $(v, k; n)$-*difference packing*, or $n$-DP$(v, k)$, is a set $\mathcal{X} = \{X_i | 1 \leq i \leq n\}$, where $X_i = \{a_{ij} | 1 \leq j \leq k\}$, for $1 \leq i \leq n$, are sets of residues modulo $v$ such that for $1 \leq i, i' \leq n, 1 \leq j \neq j' \leq k$, and $1 \leq \ell \neq \ell' \leq k$, we have $a_{ij} - a_{ij'} \equiv a_{i'\ell} - a_{i'\ell'} \pmod v$ only if $i = i', j = j'$, and $\ell = \ell'$. Difference packings and difference triangle sets are intimately related in many ways. In particular, the following observation is made by Chen, Fan, and Jin [16].

*Lemma 1 (Chen, Fan, and Jin):* An $n$-$DP(v, k)$ is an $(n, k-1)$-D$\Delta$S, whose scope is at most $v - 1$.

Furthermore, using Singer's construction [17] of 1-DP$(q^2 + q + 1, q + 1)$ for prime powers $q$ and a technique of Colbourn and Colbourn [18], they constructed two infinite families of difference packings, one of which is given below.

*Theorem 3 (Chen, Fan, and Jin):* For any prime power $q$ and prime $n > q$, there exists an $n$-DP$(n(q^2 + q + 1), q + 1)$.

It is known [1, Theorem 6] that for any fixed $k$

$$\lim_{n \to \infty} m(n, k)/nk^2$$

exists and equals one. Here, we show that the same conclusion holds even if one allows $k$ to grow with $n$, provided that it does not grow too fast. The following result of Heath-Brown and Iwaniec [19] on differences between consecutive primes is useful.

*Theorem 4 (Heath-Brown and Iwaniec):* Let $p_n$ denote the $n$th prime. Then

$$p_{n+1} - p_n \ll p_n^{11/20 + \epsilon}$$

for any $\epsilon > 0$.

*Theorem 5:* Let $n$ and $k$ be positive integers such that $n > k$ or $n = 1$. Then there exists an $(n, k)$-D$\Delta$S whose scope is at most $(1 + o(1))nk^2$, where the $o(1)$ is with respect to $k$.

*Proof:* Suppose $n > k$. Let $p$ and $q$ be the smallest prime at least $n$ and $k$, respectively, such that $p > q$. Then Theorem 3 assures us of the existence of a $p$-DP$(p(q^2 + q + 1), q + 1)$. This difference packing is a $(p, q)$-D$\Delta$S by Lemma 1. Hence, by repeated shortening and reduction (if necessary), we obtain an $(n, k)$-D$\Delta$S whose scope $m$ is upper-bounded by $p(q^2 + q + 1)$. However, Theorem 4 implies that

$$p(q^2 + q + 1) \leq (1 + o(1))nk^2. \qquad (1)$$

For $n = 1$, we use Singer's 1-DP $(q^2 + q + 1, q + 1)$ and follow the same argument above. □

*Corollary 1:* Let $k = f(n)$, where $f$ is an increasing function such that $\lim \sup_{n \to \infty} f(n)/n < 1$. Then

$$\lim_{n \to \infty} \frac{m(n, k)}{nk^2} = 1.$$

*Proof:* For $n$ large enough, we have $k < n$ and Theorem 5 can be used to give an $(n, k)$-D$\Delta$S of scope at most $(1 + o(1))nk^2$. This, together with Theorem 1 yields the desired result. □

## III. EXHAUSTIVE SEARCH

In principle, to construct an $(n, k)$-D$\Delta$S of scope $m$, we must consider all possible sets of $n$ subsets from a universe of size $\binom{m}{k}$. This is only feasible for small values of $n$ and $k$. Nevertheless, the main advantage of exhaustive search is that it allows us to prove the nonexistence of $(n, k)$-D$\Delta$S of certain scopes.

The existing results on difference triangle sets present few unknown values of $m(n, k)$ that can be determined exactly with today's technology. The determination of $m(2, 7)$ is one of these possibilities. It is known that $61 \leq m(2, 7) \leq 73$ (see [1]). We proved that $m(2, 7) = 70$ by employing a backtracking algorithm that ran for about a week on a network of 30 machines for undergraduate mathematics students at the University of Waterloo. The blocks of a $(2, 7)$-D$\Delta$S of scope 70 are given below:

$$X_1 = \{0, 1, 4, 24, 40, 54, 67, 69\}$$

and

$$X_2 = \{0, 6, 11, 18, 28, 37, 62, 70\}.$$

We did not attempt to find all $(2, 7)$-D$\Delta$S of scope 70.

In the following sections, we turn to faster heuristics for constructing difference triangle sets.

## IV. GREEDY ALGORITHMS

We define a *partial* $(n, k)$-D$\Delta$S to be a set $\mathcal{X} = \{X_i | 1 \leq i \leq s\}$ satisfying all of the following conditions:

1) $s \leq n$.
2) $|X_i| = k_i + 1 \leq k + 1$, for $1 \leq i \leq s$.
3) $X_i = \{a_{ij} | 0 \leq j \leq k_i\}$ is such that
$$0 = a_{i0} < a_{i1} < \cdots < a_{ik_i}, \qquad \text{for } 1 \leq i \leq s.$$
4) The differences $a_{ij} - a_{ij'}$, for $1 \leq i \leq s$, and $0 \leq j \neq j' \leq k_i$ are all distinct and nonzero.

The *trivial partial* $(n, k)$-D$\Delta$S is the partial $(n, k)$-D$\Delta$S $\mathcal{X} = \{X_i | 1 \leq i \leq n\}$ such that $X_i = \{0\}$ for $1 \leq i \leq n$. With the above definition, an $(n, k)$-D$\Delta$S is a partial $(n, k)$-D$\Delta$S $\mathcal{X} = \{X_i | 1 \leq i \leq s\}$, where $s = n$ and $|X_i| = k + 1$ for $1 \leq i \leq s$.

Every partial $(n, k)$-D$\Delta$S $\mathcal{X} = \{X_i | 1 \leq i \leq s\}$ has a *representation* by an $n \times (k + 1)$ array $R = (r_{ij}), 1 \leq i \leq n$ and $0 \leq j \leq k$, where each cell is either empty or contains a nonnegative integer. The entries of the nonempty cells in row $i$ of $R$ are exactly the members of $X_i$. Let $R(n, k)$ denote the $n \times (k + 1)$ array with all the cells in the zeroth column containing zeros and all other cells empty. Then $R(n, k)$ is an array representation for the trivial partial $(n, k)$-D$\Delta$S.

The greedy algorithms we propose can be conveniently described in terms of these array-representations for difference triangle sets.

Our first algorithm, called the set-greedy algorithm, works as follows. We begin with $R(n, k)$. At each stage of the algorithm, we pick the smallest $i$ such that the $i$th row contains an empty cell. We place in this empty cell the smallest positive integer such that the resulting array remains a representation of a partial $(n, k)$-D$\Delta$S. The algorithm terminates when the array contains no empty cells. The idea behind this algorithm is to fill in the empty cells of $R(n, k)$ in a row-by-row manner. This suggests the following variant which fills in the empty cells of $R(n, k)$ column-wise.

The transversal-greedy algorithm also starts with $R(n, k)$. At each stage of the algorithm, we pick the smallest $j$ such that the $j$th column contains an empty cell. We then fill in the first empty cell of this column with the smallest positive integer so that the resulting array remains a representation of a partial $(n, k)$-D$\Delta$S. The algorithm terminates when the array contains no empty cells.

It is evident that both of the above algorithms terminate with an array representation of an $(n, k)$-D$\Delta$S. Empirical computational results show that the transversal-greedy algorithm outperforms the set-greedy algorithm. There is also an interesting connection between the transversal-greedy algorithm and a certain combinatorial game introduced by Wythoff [20] in 1907.

In Wythoff's game, there are two players who play alternately. Initially, there are two piles of matches, $r$ in each pile. A player may take an arbitrary number of matches from one pile or an equal number of matches from each of the two piles, but he must take at least one match. The player who takes the last match wins the game. The *position* of a player is the pair $(u, v)$ where $u$ is the number of matches left in one pile and $v$ is the number of matches left in the other, immediately after his/her move. Without loss of generality, we assume that $u \leq v$. A player's position is *winning* if no matter what his/her opponent does, the player can force a win. Define the numbers $u_i$ and $v_i$ recursively as follows:

1) $u_1 = 1$;
2) $v_i = u_i + i$;
3) $u_{i+1}$ is the smallest positive integer distinct from the $2i$ integers $u_1, u_2, \cdots, u_i, v_1, v_2, \cdots, v_i$.

TABLE I
IMPROVED UPPER BOUNDS FOR $m(n,k)$

| k | 2 | 3 | 4 | 5 | 6 | 7 | 8 | 9 | 10 | 11 | 12 | 13 | 14 | 15 |
|---|---|---|---|---|---|---|---|---|---|---|---|---|---|---|
| 4 | | | | | | | | | | 123 [144] | 133 {159} | 146 {160} | 156 {161} | |
| 5 | | | | | 110 [111] | 130 {140} | 145 [170] | 170 [185] | 186 [213] | 204 [214] | 222 [254] | 234 [258] | 259 [288] | 275 [321] |
| 6 | | | 117 (122) | 146 (162) | 172 (192) | 198 (208) | 225 {245} | 251 [306] | 277 [312] | 314 [356] | 340 (403) | 366 (443) | 393 (496) | 432 [535] |
| 7 | | 126 (127) | 164 (169) | 206 (224) | 249 (263) | | 327 {360} | 368 {376} | 415 [425] | 438 [439] | | | | |
| 8 | 100 (102) | 163 (166) | | | 346 (353) | | | 518 {528} | | | | | 797 [834] | 845 [849] |
| 9 | | | | | | | | | | | | | 1046 [1097] | 1088 [1121] |
| 10 | | | | | | | | | | | | | 1362 (1435) | 1415 (1544) |
| 11 | | | | | | | | | | | | | | |
| 12 | | | | | | | | | | | | | | 2234 [2237] |

Connell [21] has shown that the winning positions for Wythoff's game are exactly those pairs $(u_i, v_i)$, for $i \geq 1$, together with $(0,0)$.

*Theorem 6:* Let $R = (r_{ij})$ be the array-representation of an $(n,2)$-D$\Delta$S constructed by the transversal-greedy algorithm. Then $(r_{i2} - n - i, r_{i2} - n)$, for $1 \leq i \leq n$, are winning positions for Wythoff's game.

*Proof:* We show that $r_{i2} - n - i = u_i$. The proof is by induction on $i$. It is easy to see that after the transversal-greedy algorithm fills the cells of the first column of $R(n,k)$, we have $r_{i1} = i$, for $1 \leq i \leq n$. These generates also the $n$ differences $1, 2, \cdots, n$. Hence, the smallest integer that the first cell in the second column can receive is $n+2$. Hence,

$$r_{1,2} - n - 1 = n + 2 - n - 1 = 1 = u_1.$$

Now assume that for some $j \geq 1$, the entries $r_{i2}$ filled in by the transversal-greedy algorithm satisfy $r_{i2} - n - i = u_i$ for $1 \leq i \leq j$. Then, the set of differences to be avoided is

$$D = \{1, 2, \cdots, n\} \cup \{r_{i2} | 1 \leq i \leq j\} \cup \{r_{i2} - i | 1 \leq i \leq j\}.$$

We consider how the transversal-greedy algorithm next determines $r_{j+1,2}$. Clearly, $r_{j+1,2}$ is the smallest positive integer such that
  i) $r_{j+1,2} \notin D$; and
  ii) $r_{j+1,2} - (j+1) \notin D$.
These conditions are satisfied if and only if $r_{j+1,2} - n - (j+1) \notin D'$ where

$$D' = \{r_{i2} - n | 1 \leq i \leq j\} \cup \{r_{i2} - n - i | 1 \leq i \leq j\}.$$

The induction hypothesis implies that

$$D' = \{u_i | 1 \leq i \leq j\} \cup \{v_i | 1 \leq i \leq j\}$$

and hence $r_{j+1,2} - n - (j+1) = u_{j+1}$. This completes the proof. □

*Corollary 2:* The scope of the $(n,2)$-D$\Delta$S constructed by the transversal-greedy algorithm is $\lfloor (5 + \sqrt{5})/2)n \rfloor$.

*Proof:* Follows from Connell's result [21] that

$$u_i = \lfloor (1 + \sqrt{5})i/2 \rfloor.$$

□

It follows that the $(n,2)$-D$\Delta$S constructed by the transversal-greedy algorithm is only about a factor of $1.21$ worse than the optimal. Analysis of the scope of $(n,k)$-D$\Delta$S constructed by the transversal-greedy algorithm, for any $k \geq 3$, seems difficult.

## V. RANDOMIZED HEURISTICS

We describe in this section some randomized heuristics that have been very effective in constructing difference triangle sets of small scope. These heuristics fit into a general framework. An $(n,k)$-*template* is a subset of $\{1, 2, \cdots, n\} \times \{0, 1, \cdots, k\}$. There is a natural correspondence between a set of cells of an $n \times (k+1)$ array and an $(n,k)$-template.

| Randomized Heuristic |
|---|
| **Step 1**: Let $R$ be an array representation of any $(n,k)$-D$\Delta$S. |
| **Step 2**: Let $\mathcal{T}$ be a set of $(n,k)$-templates. |
| **Step 3**: Repeat Step 4 to Step 6 $N$ times: |
| **Step 4**: Let $s$ be the scope of the $(n,k)$-D$\Delta$S represented by $R$. |
| **Step 5**: Randomly choose $(n,k)$-template in $\mathcal{T}$ and delete the entries in the cells of $R$ corresponding to that template. |
| **Step 6**: Find all possible ways of filling in the empty cells of $R$ using nonnegative integers no greater than $s$. Randomly choose one of these ways and fill in the empty cells of $R$ accordingly. |

The final difference triangle set constructed by the randomized heuristic have scope at most that of the initial difference triangle set, and strictly improves on the scope if at any stage of the algorithm, the cell containing the largest element of the array is emptied and this element is never used to refill any cell.

The sets of templates $\mathcal{T}$ that we find most effective in constructing difference triangle sets of small scope are the three listed below.
  1) $\mathcal{T}_1 = \{\{(i,j)\} | 1 \leq i \leq n, 0 \leq j \leq k\}$;
  2) $\mathcal{T}_2 = \{\{(i,j) | 0 \leq j \leq k\} | 1 \leq i \leq n\}$;
  3) $\mathcal{T}_3 = \{\{(i,j_i) | 1 \leq i \leq n\} | 0 \leq j_i \leq k\}$.

The templates $\mathcal{T}_1, \mathcal{T}_2$, and $\mathcal{T}_3$ correspond to emptying single cells, emptying the cells in an entire row, and emptying one cell from each row in Step 5 of the randomized heuristic, respectively. Let us denote by $H_i$ the randomized heuristic that uses $\mathcal{T}_i$. Naturally, $H_1$ is the fastest. $H_2$ and $H_3$ are much slower but generally give better results when $n$ and $k$ are small. $H_2$ is effective when $n/k$ is large whereas $H_3$ is effective when $n/k$ is small. These heuristics work well when the initial array-representation of an $(n,k)$-D$\Delta$S used in Step 1 is that constructed by the transversal-greedy algorithm. We used these three heuristics in combination, typically in the order $H_1, H_2, H_3$ or $H_1, H_3, H_2$, to obtain a number of improvements on the upper bounds for $m(n,k)$. Instances of these improvements are given in Table I, where our improved upper bound is given above the best previous upper bound. The bounds in parentheses "( )" are due to Kløve [1], those in brackets "[ ]" are due to Chen, Fan, and Jin [16], and those in braces "{ }" are due to Chen [15]. The blocks for difference triangle sets with the improved scopes are available from the authors.

## VI. CONCLUDING REMARKS

One of the problems suggested by the results in Section II is the determination of the asymptotic behavior of $m(n,k)$. Our results show that for $f(n)$ satisfying $\lim \sup_{n\to\infty} f(n)/n < 1$, we have $\lim_{n\to\infty} m(n,f(n))/n(f(n))^2 = 1$. It would be interesting to know what happens if $f(n)$ is allowed to grow at a faster rate.

We have also described algorithms that are used to construct difference triangle sets with the best known scopes for many intermediate values of $n$ and $k$.

# Contribution to Munuera's Problem on the Main Conjecture of Geometric Hyperelliptic MDS Codes

Hao Chen and Stephen S.-T. Yau, *Senior Member, IEEE*

*Abstract*—In coding theory, it is of great intererst to know the maximal length of MDS codes. In fact, the Main Conjecture says that the length of MDS codes over $F_q$ is less than or equal to $q+1$ (except for some special cases). Munuera proposed a new way to attack the Main Conjecture on MDS codes for geometric codes. In particular, he proved the conjecture for codes arising from curves of genus one or two when the cardinal of the ground field is large enough. He also asked whether a similar theorem can be proved for any hyperelliptic curve. The purpose of this correspondence is to give an affirmative answer. In fact, our method also proves the Main Conjecture for geometric MDS codes for $q = 2$ if the genus of the hyperelliptic curve is either $1$, $2$ or $3$, and for $q = 3$ if the genus of the curve is $1$.

*Index Terms*— Algebraic curves, algebraic-geometric codes, divisors, hyperelliptic curves, zeta function.

## I. INTRODUCTION

Let $F_q$ be a finite field with $q$ elements and $X$ be a nonsingular projective curve defined over $F_q$ with genus $g$. We shall write $X(F_q)$ to indicate the finite set of $F_q$-rational points on $X$. The function field of $X$ over $F_q$ is denoted by $F_q(X)$. Let $\mathcal{P} = \{P_1, \cdots, P_n\}$ be a set of $n$ distinct rational points on $X$. By abusing notation, we also sometimes identify $\mathcal{P}$ as a divisor. Let $G$ be a rational divisor with support disjoint from $\mathcal{P}$.

$$L(G) := \{f \in F_q(X) : (f) + G \geq 0\} \cup \{0\} = H^0(X, [G])$$

where $[G]$, the line bundle corresponding to the divisor $G$, is a vector space, and we denote $\ell(G)$ its dimension. The complete linear system associated to $G$, denoted by $|G|$, is

$$\{f \in F_q(X) : (f) + G \geq 0\}/F_q^*.$$

*Definition:* The algebraic geometry code $C(X, \mathcal{P}, G)$ associated to the pair $(\mathcal{P}, G)$ is the linear code of length $n$ defined as the image of the linear map

$$\begin{aligned} \alpha : L(G) &\to F_q^n \\ f &\to (f(P_1), \cdots, f(P_n)). \end{aligned}$$

We shall let $k$ denote the dimension of this linear code. Then $k = \ell(G) - \ell(G - \mathcal{P})$. In what follows, we shall always assume that

$$2g - 2 < \deg G < n. \tag{1.1}$$

It is well known that the dimension $k$ and the minimum distance $d$ of the algebraic geometric code $C(X, \mathcal{P}, G)$ satisfy the following relations [11]:

$$k = \ell(G) = \deg G + 1 - g \tag{1.2}$$

$$d \geq n - \deg G. \tag{1.3}$$

Manuscript received July 12, 1996; revised February 3, 1997. The work of H. Chen was supported by the NSF of China and by the Guangdong Provincial NSF of China. The work of S. S.-T. Yau was supported in part by ARO DAAH04-1-0530 and NSF DMS 9321262.
H. Chen is with the Department of Mathematics, Zhongshan University, Guangzhou, Guangdong 510275, P.R. China.
S. S.-T. Yau is with the Department of Mathematics, Statistics, and Computer Science, University of Illinois at Chicago, Chicago, IL 60607-7045 USA.
Publisher Item Identifier S 0018-9448(97)03868-6.